\documentclass[aps,amssymb,superscriptaddress]{revtex4}

\usepackage{epsfig, color}
\usepackage{amsthm, amsmath, amsfonts, ae, psfrag, bbold}

\newcommand{\upd}{\mathrm{d}}
\newcommand{\bvec}[1]{{\bf\string#1 }}

\begin{document}

\title{A fundamental measure theory for the sticky hard sphere fluid}

\author{Hendrik Hansen-Goos}
 \email{hendrik.hansen-goos@yale.edu}

\affiliation{Department of Geology and Geophysics, Yale University, New Haven, CT 06520, USA
}

\author{J.~S.~Wettlaufer}

\affiliation{Department of Geology and Geophysics, Yale University, New Haven, CT 06520, USA
}

\affiliation{Department of Physics, Yale University, New Haven, CT 06520, USA}

\affiliation{Program in Applied Mathematics, Yale University, New Haven, CT 06520, USA}

\date{\today}   

\begin{abstract}

We construct a density functional theory (DFT) for the sticky hard sphere (SHS) fluid which, like Rosenfeld's fundamental measure theory (FMT) for the hard sphere fluid [Phys. Rev. Lett. {\bf 63}, 980 (1989)], is based on a set of weighted densities and an exact result from scaled particle theory (SPT). It is demonstrated that the excess free energy density of the inhomogeneous SHS fluid $\Phi_{\text{SHS}}$ is uniquely defined when (a) it is solely a function of the weighted densities from Kierlik and Rosinberg's version of FMT [Phys. Rev. A {\bf 42}, 3382 (1990)], (b) it satisfies the SPT differential equation, and (c) it yields any given direct correlation function (DCF) from the class of generalized Percus-Yevick closures introduced by Gazzillo and Giacometti [J. Chem. Phys. {\bf 120}, 4742 (2004)]. The resulting DFT is shown to be in very good agreement with simulation data. In particular, this FMT yields the correct contact value of the density profiles with no adjustable parameters. Rather than requiring higher order DCFs, such as perturbative DFTs, our SHS FMT produces them. Interestingly, although equivalent to Kierlik and Rosinberg's FMT in the case of hard spheres, the set of weighted densities used for Rosenfeld's original FMT is insufficient for constructing a DFT which yields the SHS DCF.

\end{abstract}

\maketitle

\section{Introduction}

Density functional theory (DFT) was developed to characterize quantum systems in an external potential \cite{HoKo64} and since the 1970's has been fruitfully brought to bear on classical fluids \cite{Ev79}. Indeed, here too it serves as an excellent tool to study the microscopic structure of fluids under the influence of external potentials, such as confining walls, and can serve as the sole prerequisite to map out complete phase diagrams that encompass various fluid and crystalline structures. Moreover, besides yielding a sound theoretical understanding of findings from colloidal physics and computer simulations, DFT calculations can be extremely efficient computationally, especially if the systems under consideration allow for dimensional reduction, and hence in consequence largely outperform Monte Carlo (MC) simulations. In particular, DFT is especially convenient when phase diagrams or solvation free energies are calculated because it makes direct use of the grand potential $\Omega$. The basic formalism of DFT expresses $\Omega$ as a functional of the density profiles $\rho_i(\bvec{r})$ of the different fluid components $i = 1,\ldots,\nu$ as
\begin{equation}
\label{eq_defOmega}
  \Omega[\{\rho_i(\bvec{r})\}]  = \mathcal{F}_{\text{id}}[\{\rho_i(\bvec{r})\}] + \mathcal{F}_{\text{ex}}[\{\rho_i(\bvec{r})\}]  + \sum_{i=1}^{\nu} \int \upd \bvec{r} \rho_i(\bvec{r}) [ V_i^{\text{ext}}(\bvec{r}) - \mu_i ] \, ,
\end{equation}
where $\mu_i$ is the chemical potential of species $i$, and $V_i^{\text{ext}}(\bvec{r})$ the external potential acting on particles of species $i$. The ideal gas part of the free energy is
\begin{equation}
  \beta \mathcal{F}_{\text{id}}[\{\rho_i(\bvec{r})\}] = \sum_{i=1}^{\nu} \int \upd \bvec{r} \rho_i(\bvec{r}) ( \ln [\Lambda_i^3\rho_i(\bvec{r} ) ] - 1) \, ,
\end{equation}
where $\Lambda_i$ is the thermal wavelength of species $i$ and $\beta = (k_B T)^{-1}$. The excess part of the free energy $\mathcal{F}_{\text{ex}}[\{\rho_i(\bvec{r})\}]$ depends on the interactions between the particles and on temperature. For the given external potentials, the equilibrium density profiles are obtained by solving the system of coupled equations $\delta \Omega/\delta \rho_i \equiv 0$, $i = 1,\ldots,\nu$.

The above formalism is exact subject to the limitations of knowledge regarding $\mathcal{F}_{\text{ex}}$ for which we have only approximate descriptions for 3D fluids with non-trivial interactions. Among these approximations, Rosenfeld's fundamental measure theory (FMT) for the hard-sphere (HS) fluid \cite{Ro89} provides an excellent framework that uses only exact results for certain limiting cases which are generalized using dimensional analysis of a set of weighted densities. The theory applies equally to mixtures and, in part thanks to subsequent incorporations of highly accurate equations of state \cite{RoEvetal02, HaRo06}, compares extremely well with computer simulations. The theory can be extended to treat crystallization of the HS fluid \cite{RoSchmLoeTa97, TaRo97, Ta00} as well as the isotropic-nematic transition of non-spherical hard particles \cite{Ro94, Ro95, HaMe09, HaMe10}. All of these developments make FMT a central and powerful tool for the study of entropy-governed hard-body fluids.

Hard-body systems are an important test bed for theories which make use of a Weeks-Chandler-Andersen \cite{WeChaAn71} type decomposition of the interaction potential into a repulsive core and an attractive long-ranged tail. Moreover, they can often be considered as an accurate approximation to an entropy-dominated system. However, indeed many realistic fluids that occur in biological applications or colloidal science invariably exhibit attractive particle-particle interactions in addition to the hard-core repulsion. A simple model fluid with an attractive potential that has been extensively studied by theorists \cite{BaHe67, SPBeCa05} is the square-well (SW) fluid defined by an interaction potential
\begin{equation}
  \beta \phi_{\text{SW}}(r) = 
  \begin{cases}
     \infty\, , \qquad 0 \le r < \sigma \\
     -\beta \epsilon_{\text{SW}} \, ,\qquad \sigma \le r \le \sigma' \\
     0 \, ,\qquad r>\sigma'
  \end{cases}
\end{equation}
where $\sigma= 2 R$ is the particle hard-core diameter, $\sigma'-\sigma$ is the range of the attraction and $\epsilon_{\text{SW}}$ is the well depth.

The radial distribution function (RDF) $g(r)$ and the direct correlation function (DCF) $c(r)$ are related through the Ornstein-Zernike (OZ) integral equation \cite{HaMD86}
\begin{equation}
\label{eq_OZ}
  g(r) - 1 =  c(r) + \rho \int \upd \bvec{r}' c(r') (g(|\bvec{r}-\bvec{r}'|) - 1) \, .
\end{equation}
However, the OZ equation can only be solved numerically within the Percus-Yevick (PY) closure for the potential $\phi_{\text{SW}}$ \cite{SmHeMu74}. Alternatively, $g(r)$ can be determined in computer simulations \cite{LaSoYuSa05} from which $c(r)$ and thermodynamic properties of the SW fluid can be derived. However, these results cover only a limited selection of specific parameter choices for the model simulations and the results are clearly not in a convenient form for analysis. Hence, it is reasonable to seek a model with attractive interactions that would be solvable within the usual closure relations for the OZ equation. In the limiting case of surface adhesion where
\begin{equation}
  \sigma'-\sigma \to 0 \qquad \text{and} \qquad \beta \epsilon_{\text{SW}} = \ln \frac{\sigma}{12\tau (\sigma'-\sigma ) } \to \infty \, ,
\end{equation}
Baxter's so-called sticky limit, with $\tau^{-1}$ characterizing the strength of the attraction, allows the OZ equation to be solved within the PY closure \cite{Ba68}. The results are convenient in a closed analytical form and compare well with computer simulations for various attraction strengths $\tau^{-1}$ \cite{MiFr04gr}. Essential features of the sticky hard sphere (SHS) fluid, such as the existence of a critical point associated with a liquid-gas transition, are captured by the PY results \cite{MiFr04pd}. Subsequently, the model has been generalized for SHS mixtures where the PY closure leads to a set of coupled quadratic equations that can be solved analytically only in special cases \cite{PeSm75, CuPeSm76}.

Far from being just a somewhat artificial theoretical construct, the SHS fluid has been demonstrated to provide an accurate description of colloidal systems with very short ranged attractions \cite{ChRoScTa94, VeDh95, BuRuPi07} and has been applied successfully to phenomena such as structural arrest in a micellar system \cite{Maetal00}, competitive protein adsorption \cite{Di92}, and even skim milk acidification \cite{Kr97}. That said, the use of the SHS fluid to model experimental systems has been aimed at mapping out phase diagrams and calculating equations of state while the inhomogeneous fluid has been studied less thoroughly. Firstly, this has to do with the difficulty of measuring structural properties such as density profiles and higher order correlation functions. Secondly, we lack a reliable tool to compute these quantities based on the analytical PY results. Existing DFTs for Baxter's SHS fluid are all based on an expansion of the one-particle DCF in powers of the deviation of the density profile from the bulk density. This expansion can only be performed systematically to linear order where it involves the DCF for which the PY results can be used \cite{ChGh97}. Moreover, the quality of the resulting DFT is rather poor, especially close to contact with a hard wall \cite{ChGh97}. Expanding the one-particle DCF to quadratic order requires a formula for the unknown three-particle DCF $c^{(3)}$ of the SHS fluid. The functional form of $c^{(3)}$ has been estimated by different authors based on assumptions regarding its range \cite{HwKi98, YoKi98} or its factorization in terms of $c(r)$ \cite{Zh01}. In order to yield reasonable agreement with computer simulations all these extensions of the DFT by Choudhury and Ghosh \cite{ChGh97} require an additional, density dependent, fitting parameter which is chosen such that the correct contact value of the density profile or the correct bulk pressure is guaranteed.

Here we develop a DFT for the SHS fluid that is distinct from the above perturbative approaches. In the spirit of FMT we use a set of weighted densities that can be identified from a systematic inspection of the HS fluid in the low density limit. In order to construct a functional which yields the PY DCF for the SHS fluid, we supplement this with an exact relation from scaled particle theory (SPT) \cite{Reiss60, Reiss77}. The main result is a unique construction for a broad class of closure schemes of which the PY closure is but one. No additional input and no fitting parameters are used. Rather than {\em requiring} higher order DCFs our theory {\em produces} them.

In Sec.~\ref{sec_genPY} we present the generalized PY closure as introduced by Gazzillo and Giacometti \cite{GaGi04}. We then summarize in Sec.~\ref{sec_introFMT} the FMT for the HS fluid in the version developed by Kierlik and Rosinberg (KR) \cite{KiRo90, KiRo93}. In Sec.~\ref{sec_FMTSHS} we construct the new FMT for the single component SHS fluid, and then compare the results from the new functional with data from numerical simulations in Sec.~\ref{sec_simulations}. In the concluding section we place the new functional in a broader context, discussing both its merits and limitations with a special focus on future developments that might be pursued in the framework of our approach.

\section{Generalized Percus-Yevick approximation}

\label{sec_genPY}

Gazzillo and Giacometti introduced the generalized Percus-Yevick (GPY) approximation for the SHS fluid \cite{GaGi04}. The GPY is defined by a closure relation for the DCF $c(r)$ of the form
\begin{equation}
\label{eq_cGPY}
  c_{\text{GPY}}(r) = 
  \begin{cases}
   -[1+\gamma(r)] \, , \quad 0<r<\sigma\, , \\
  c_{\text{shrink}}(r) \, , \quad \sigma < r < \sigma' \, , \\
  0 \, , \quad r > \sigma' \, .
  \end{cases}
\end{equation}
Here $\gamma(r) = \rho\int \upd\bvec{r}' c(r') h(|\bvec{r}-\bvec{r}'|)$ is the convolution of the DCF with the total correlation function $h(r) = g(r)-1$, $\rho$ is the particle number density of the bulk fluid. Depending on the choice of the function $c_{\text{shrink}}(r)$, when applied to the SHS fluid the GPY reduces to different familiar closure relations such as the mean spherical approximation (MSA), the modified mean spherical approximation (mMSA), and the PY closure \cite{GaGi04}. For example, choosing  $c_{\text{shrink}}(r) = f(r)[1+\gamma(r)]$ in Eq.~\eqref{eq_cGPY}, where $f(r)$ is the Mayer $f$-function, results in the PY closure relation $c_{\text{PY}}(r) = f(r)[1+\gamma(r)]$ for arbitrary $r>0$.

Gazzillo and Giacometti \cite{GaGi04} showed that in Baxter's sticky limit the GPY approximation allows for a general analytical solution of the OZ equation where the underlying closure relation enters solely through its approximation to the cavity function $y_{\sigma}$ at contact. The resulting DCF is
\begin{equation}
\label{eq_GPYc}
  c_{\text{GPY}}(r) = \left\{-12\eta q_{\sigma}^2 \frac{\sigma}{r} - a^2 + 12\eta\left[\frac{1}{2}(a+b)^2-a q_{\sigma}\right]\frac{r}{\sigma} - \frac{\eta}{2} a^2 \left( \frac{r}{\sigma}\right)^3 \right\} \Theta(\sigma-r) + q_{\sigma}\sigma\delta(r-\sigma)\, ,
\end{equation}
where
\begin{equation}
  \label{eq_defa}
  a = \frac{1+2\eta}{(1-\eta)^2} - \frac{12 q_{\sigma} \eta}{1-\eta} \, , \quad
  b = -\frac{3\eta}{2(1-\eta)^2}+\frac{6q_{\sigma}\eta}{1-\eta} \, , \quad \text{and} \quad
  q_{\sigma} = \frac{y_{\sigma}}{12\tau} \, . 
\end{equation}

The PY closure corresponds to taking
\begin{equation}
\label{eq_ysigmaPY}
 y_{\sigma}^{\text{PY}}(\eta) = \frac{2 y_{\sigma}^{\text{HS}}(\eta)}{1+\frac{\eta}{(1-\eta)\tau} + \sqrt{\left(1+\frac{\eta}{(1-\eta)\tau} \right)^2 - \frac{\eta y_{\sigma}^{\text{HS}}(\eta) }{3\tau^2}} } \, ,
\end{equation}
in which
\begin{equation}
 y_{\sigma}^{\text{HS}}(\eta) = \frac{1+\eta/2}{(1-\eta)^2}
\end{equation}
is the PY result for the HS fluid. The PY result for $y_{\sigma}$ [Eq.~\eqref{eq_ysigmaPY}] is exact up to linear order in $\eta$, with a small deviation occurring at quadratic and higher orders.

It turns out that the MSA leads to $y_{\sigma}^{\text{MSA}} = 0$ and hence does not capture the effect of surface adhesion in the SHS fluid and thus the results coincide with those for the HS fluid. The mMSA has $y_{\sigma}^{\text{mMSA}} = 1$ which is the exact result in the low density limit; $\eta\to 0$. Gazzillo and Giacometti \cite{GaGi04} also introduce a C1 approximation wherein $y_{\sigma}^{\text{C1}} = 1 + \left(\frac{5}{2}-\frac{1}{\tau}+\frac{1}{12\tau^2}\right)\eta$, and hence it corresponds to the exact expression for $y_{\sigma}$ truncated after the linear term in $\eta$.

\section{Fundamental measure theory for the hard sphere fluid}

\label{sec_introFMT}

Presently, FMT provides the most accurate free energy model for inhomogeneous HS fluids. Additionally, FMT gives an excellent description of HS mixtures \cite{RoDi00} as well as fluids containing non-spherical hard particles \cite{Ro94,Ro95,HaMe09,HaMe10}. The theory originated with Rosenfeld's seminal paper \cite{Ro89} and has been continuously modified by improving the underlying equation of state \cite{RoEvetal02, HaRo06} and the ability to capture the formation of HS crystals \cite{RoSchmLoeTa97, TaRo97, Ta00}. There is an equivalent FMT formulation due to Kierlik and Rosinberg \cite{KiRo90, KiRo93}, which, in the case of the SHS fluid proves to be more useful than Rosenfeld's FMT (cf. Sec.~\ref{sec_conclusion}). In order to make this development reasonably self-contained, we summarize KR's version of FMT here.

Fundamental measure theory provides an expression for the excess free energy $\mathcal{F}_{\text{ex}}$ which can be used in the framework of classical DFT to obtain the structural properties (density profiles and correlation functions) of fluids that are subject to external potentials, i.e., inhomogeneous fluids. As discussed above, we consider a fluid with $\nu$ different components which have the corresponding density profiles $\rho_i(\bvec{r})$, $i = 1, \ldots , \nu$. The FMT approach is based on weighted densities $n_{\alpha}(\bvec{r})$ written as
\begin{equation}
\label{eq_defwd}
  n_{\alpha}(\bvec{r}) = \sum_{i=1}^{\nu} \int \upd \bvec{r}' \rho_i(\bvec{r}') \omega_i^{(\alpha)}(\bvec{r}-\bvec{r}')
\end{equation}
in terms of the weight functions $\omega_i^{(\alpha)}$ of which there are four in KR's approach corresponding to four $n_{\alpha}$. These can be described in Fourier space:
\begin{align}
\label{eq_omega3}
   \omega_i^{(3)}(k) & = 4\pi[\sin(k R_i)-k R_i \cos(k R_i)]/k^3 \, , \\
   \omega_i^{(2)}(k) & = 4\pi R_i \sin(k R_i)/k \, ,\\
\label{eq_omega1}
   \omega_i^{(1)}(k) & = [\sin(k R_i) + k R_i \cos(k R_i)]/2k \, , \quad \text{and} \\
\label{eq_omega0}
   \omega_i^{(0)}(k) & = \cos(k R_i) + (k R_i/2) \sin(k R_i) \, ,
\end{align}
or in real space as
\begin{align}
\label{eq_omega3real}
   \omega_i^{(3)}(r) & = \Theta(R_i-r) \, , \\
   \omega_i^{(2)}(r) & = \delta(R_i-r) \, , \\
   \omega_i^{(1)}(r) & =  \delta'(R_i-r)/8\pi \, , \quad \text{and} \\
\label{eq_omega0real}
   \omega_i^{(0)}(r) & =  -\delta''(R_i-r)/8\pi +  \delta'(R_i-r)/2\pi R_i - \delta(R_i-r)/2\pi R_i^2 \, .
\end{align}
Here, $\Theta$ is the Heaviside function and the primes denote differentiation. Using these weighted densities, the FMT excess free energy for the hard sphere mixture is readily obtained as
\begin{equation}
  \beta \mathcal{F}_{\text{ex}}[\{\rho_i\}] = \int \upd\bvec{r} \Phi_{\text{HS}}(\{n_{\alpha}(\bvec{r})\}) \, ,
\end{equation}
where the excess free energy density $\Phi_{\text{HS}}$ is given by
\begin{equation}
\Phi_{\text{HS}} = -n_0 \ln(1-n_3) + \frac{n_1 n_2}{1-n_3} + \frac{n_2^3}{24\pi(1-n_3)^2} \, .
\end{equation}
Fundamental measure theory, which may be derived independently of PY integral theory, can be shown to yield the same equation of state and the same DCF as obtained using the PY closure.

\section{Fundamental measure theory for the sticky hard sphere fluid}

\label{sec_FMTSHS}

The FMT excess free energy density $\Phi_{\text{HS}}$ obeys the SPT differential equation
\begin{equation}
\label{eq_SPTdiff}
  \frac{\partial \Phi}{\partial n_3} = n_0 - \Phi + \sum_{\alpha} n_{\alpha} \frac{\partial \Phi}{\partial n_{\alpha}} \, ,
\end{equation}
which insures that the reversible work of adding a sphere to the fluid, in the limit of infinite sphere size, equals the volume of the sphere multiplied by the pressure of the bulk fluid \cite{Reiss60, Reiss77}. While this exact relation must hold for spheres with surface adhesion, there is no straightforward generalization of the principle for fluids with a finite range of interaction (in addition to the hard body repulsion) such as in the case of a SW fluid. For these fluids the question arises as to how the range of the interaction should scale as the size of the added sphere goes to infinity. However, for the SHS fluid the surface attraction is subdominant with respect to the sphere volume in the large sphere limit. Therefore, Eq.~\eqref{eq_SPTdiff} must be obeyed.

Requiring Eq.~\eqref{eq_SPTdiff} to hold for the SHS fluid of radius $R$ and Baxter parameter $\tau$ implies that the corresponding free energy density $\Phi_{\text{SHS}}$ must have the form
\begin{equation}
\label{eq_defPhiSHS}
  \Phi_{\text{SHS}} = \Phi_{\text{HS}} + R^{-1}n_1 \,\phi_1\!\left(\frac{R n_2}{1-n_3}\right) + \frac{R^{-2} n_2}{2\pi} \,\phi_2\!\left(\frac{R n_2}{1-n_3}\right) \, ,
\end{equation}
where $\phi_1$ and $\phi_2$ are arbitrary dimensionless functions of the dimensionless argument. We have made the additional assumption that $\Phi_{\text{SHS}}$ is linear in $n_1$ which follows from inspection of the DCF $c_{\text{FMT}}$ obtained from an excess free energy density $\Phi$ that is solely a function of the weighted densities $n_{\alpha}$. Then, in Fourier space we have 
\begin{equation}
\label{eq_cFMT}
  c_{\text{FMT}}(k) = -\sum_{\alpha, \beta} \frac{\partial^2 \Phi}{\partial n_{\alpha} \partial n_{\beta} } \omega^{(\alpha)}(k) \omega^{(\beta)}(k) \, ,
\end{equation}
where, because we will treat only the single component fluid, we have dropped the second index of the weight functions. From the definition of the weight functions [Eqs.~\eqref{eq_omega3} -- \eqref{eq_omega0}] it can be seen that the inverse Fourier transform of $\omega^{(1)}(k) \omega^{(1)}(k)$ does not converge and hence, to insure that the DCF is defined in real space, we require $\Phi_{\text{SHS}}$ to be linear in $n_1$. A similar argument leads us to require $\Phi_{\text{SHS}}-\Phi_{\text{HS}}$ to be independent of $n_0$.

A crucial observation is that for any closure of the GPY class, there is a unique choice for the functions $\phi_1$ and $\phi_2$ in Eq.~\eqref{eq_defPhiSHS} such that $c_{\text{FMT}}$ [Eq.~\eqref{eq_cFMT} using Eq.~\eqref{eq_defPhiSHS}] equals $c_{\text{GPY}}$ from Eq.~\eqref{eq_GPYc}. This can be shown by comparing $c_{\text{FMT}}$ and $c_{\text{GPY}}$ in Fourier space.

The functions $\phi_1$ and $\phi_2$ are obtained through integration of the cavity function at contact $y_{\sigma}$ associated with the given GPY closure. This is accomplished using the equations
\begin{align}
\label{eq_diffPhi1}
  \phi_1'(x)  & = - 2 \tilde{y}_{\sigma}(x)/x \, , \\
\label{eq_diffPhi2}
  [ x^2 \phi_2'(x) ]' & = \frac{\tilde{y}_{\sigma}(x)^2}{2} - x \tilde{y}_{\sigma}(x)  + x \tilde{y}_{\sigma}'(x) \, ,
\end{align}
where $x = \frac{R n_2}{1-n_3}$ and hence in the bulk fluid, where $n_3 = \eta$ and $n_2 = 3 \eta / R$, we have $x = \frac{3\eta}{1-\eta}$ or $\eta = \frac{x}{3+x}$. The latter equation is used to rewrite $\tilde{y}_{\sigma}=\eta y_{\sigma}/\tau$ as a function of $x$ rather than as a function of $\eta$. The constants associated with the solution of Eqs.~\eqref{eq_diffPhi1} and \eqref{eq_diffPhi2} are determined from the condition that $\phi_1$ and $\phi_2$ must vanish as $x\to 0$.

The PY closure is achieved by using  $y_{\sigma}^{\text{PY}}$ from Eq.~\eqref{eq_ysigmaPY} in the formulas above and we find
\begin{align}
\label{eq_phi1PY}
 \phi_1^{\text{PY}}(x) = & 12 \tau \left(\sqrt{X}-1\right) - 4 x - 12\tau\ln Y+ 2\sqrt{2}\left(6\tau-1 \right) \ln Z \, , \quad \text{and} \\
\label{eq_phi2PY}
 \phi_2^{\text{PY}}(x)  = & \frac{36\tau^3}{x}\left(\sqrt{X}-1\right) - 6\tau^2\left(5\sqrt{X}-3\right) + \tau\left(3+9 x -\sqrt{X}-x\sqrt{X}\right) + \frac{x^2}{6} \nonumber \\
  & + 36 \tau^2 \ln Y + \frac{1}{\sqrt{2}}\left( 1 - 54 \tau^2 \right) \ln Z \, ,
\end{align}
where
\begin{align}
  X & = \left(1+\frac{x}{3\tau} \right)^2 - \frac{x}{9\tau^2}\left(1+\frac{x}{2}\right) \, , \\
  Y & = \frac{1}{2}\left(\sqrt{X}+1\right)+\frac{x}{6\tau}-\frac{x}{36\tau^2} \, , \quad \text{and} \\
  Z & = \frac{\frac{x-1}{3\tau}+2+\sqrt{2X} }{2+\sqrt{2}-\frac{1}{3\tau} } \, .
\end{align}

The MSA for the SHS fluid has $y_{\sigma}^{\text{MSA}}=0$, and hence $\phi_1=\phi_2=0$, giving identical results to those for the HS fluid without surface adhesion. The mMSA and the C1 approximation can be summarized by using the low density limit expression $y_{\sigma}^{\text{LDL}}=1+y_1 \eta$ with $y_1=0$ to recover mMSA and $y_1 = \frac{5}{2}-\frac{1}{\tau}+\frac{1}{12\tau^2}$ for the C1 approximation.  Solving Eqs.~\eqref{eq_diffPhi1} and \eqref{eq_diffPhi2} with $y_{\sigma}^{\text{LDL}}$ yields the functions
\begin{align}
  \phi_1^{\text{LDL}}(x) = & \frac{2}{\tau} \left[ \frac{y_1 x}{3+x} - (1+y_1) \ln\left(1+\frac{x}{3}\right) \right] \, , \quad \text{and} \\
  \phi_2^{\text{LDL}}(x) = &  \frac{1}{2\tau^2 x}\left[ 6+x+6\tau (2+x) + 2 y_1 \left(9+x+3\tau(7+2x) \right) + y_1^2 (12+x)\right] \ln\left(1+\frac{x}{3}\right) \nonumber \\
& -\frac{1}{2\tau^2} \left( 2 + 4\tau + \tau x \right) - \frac{y_1}{2\tau^2(3+x)}\left( 18+42\tau+5x+19 \tau x+\tau x^2\right) \nonumber \\
& - \frac{y_1^2}{12\tau^2(3+x)^2}\left(216+126 x + 17 x^2 \right) \, .
\end{align}

The pressure $p_{\text{SHS}}$ related to $\Phi_{\text{SHS}}$ [Eq.~\eqref{eq_defPhiSHS}] can be computed using either side of Eq.~\eqref{eq_SPTdiff}, viz.,
\begin{equation}
  \beta p_{\text{SHS}} =   n_0 - \Phi_{\text{SHS}} + \sum_{\alpha} n_{\alpha} \frac{\partial \Phi_{\text{SHS}}}{\partial n_{\alpha}} =\frac{\partial \Phi_{\text{SHS}}}{\partial n_3} \, .
\end{equation}
With the advent of Eqs.~\eqref{eq_diffPhi1} and \eqref{eq_diffPhi2} it can be easily shown that $p_{\text{SHS}}$ obeys the compressibility equation, which for the GPY closure is
\begin{equation}
 \left(  \frac{\partial \beta p}{\partial \rho } \right)_T = a^2 \, ,
\end{equation}
where $a$ is given by the definition in Eq.~\eqref{eq_defa} [for details see Ref.~\onlinecite{GaGi04}]. Therefore, the pressure underlying the FMT is always identical to the expression obtained from the GPY closure via the compressibility route. The quality of the mMSA, C1, and PY pressures is discussed in Ref.~\onlinecite{GaGi04} where the results are compared with computer simulations.

\section{Comparison with simulation data}

\label{sec_simulations}

\subsection{Results for the radial distribution function}

As a first test for the consistency of the density functional $\Omega$, which uses the FMT excess free energy $\beta \mathcal{F}_{\text{ex}}[\{\rho_i\}] = \int \upd\bvec{r} \Phi_{\text{SHS}}(\{n_{\alpha}(\bvec{r})\})$ with $\Phi_{\text{SHS}}$ from Eq.~\eqref{eq_defPhiSHS} and the formulae $\phi_1^{\text{PY}}$ and $\phi_2^{\text{PY}}$ from Eqs.~\eqref{eq_phi1PY} and \eqref{eq_phi2PY}, we minimize $\Omega$ in a spherical geometry. A center particle identical to the fluid particles is realized via an external potential and hence the resulting density profile corresponds to the RDF $g(r)$ of the fluid up to a factor $\rho$. On the other hand, $g(r)$ can be obtained directly from the DCF $c_{\text{PY}}$ [Eq.~\eqref{eq_GPYc} with Eq.~\eqref{eq_ysigmaPY}], using the OZ equation [Eq.~\eqref{eq_OZ}]. Obviously, for an exact DCF and an exact functional the two results must be equal, but given that approximations are involved on both sides a deviation is to be expected. The degree of consistency and the agreement with numerical simulations help evaluate the quality of the FMT constructed here.

For completeness, we include the formulae for the weighted densities in spherical geometry as they follow from Eqs.~\eqref{eq_omega3} to \eqref{eq_omega0}. The result is
\begin{align}
  \label{eq_n3spherical}
  n_3(r)  & =  \frac{\pi}{r} \int_{r-R}^{r+R} \upd r' r' \rho(r') \left[R^2-(r-r')^2 \right] \\
  n_2(r)  & = \frac{2\pi R}{r} \int_{r-R}^{r+R} \upd r' r' \rho(r') \\
  n_1(r)  & = \frac{1}{4 r} \int_{r-R}^{r+R} \upd r' r' \rho(r') + \frac{R}{4 r} \left[ (r+R)\rho(r+R)+(r-R)\rho(r-R) \right] \\
  n_0(r)  & = \frac{1}{2r} \left[ (r+R/2) \rho(r+R) + (r-R/2) \rho(r-R) \right] \nonumber \\
  &  \quad - \frac{R}{4r} \left[ (r+R) \rho'(r+R) - (r-R) \rho'(r-R) \right] \, ,
  \label{eq_n0spherical}
\end{align}
where $r$, the distance from the center, is larger than $R$. Hence the problem of minimizing $\Omega$ in spherical geometry can be reduced to a computation in 1D.

\begin{figure}[t]

\begin{center}

\includegraphics[height=.33\textwidth]{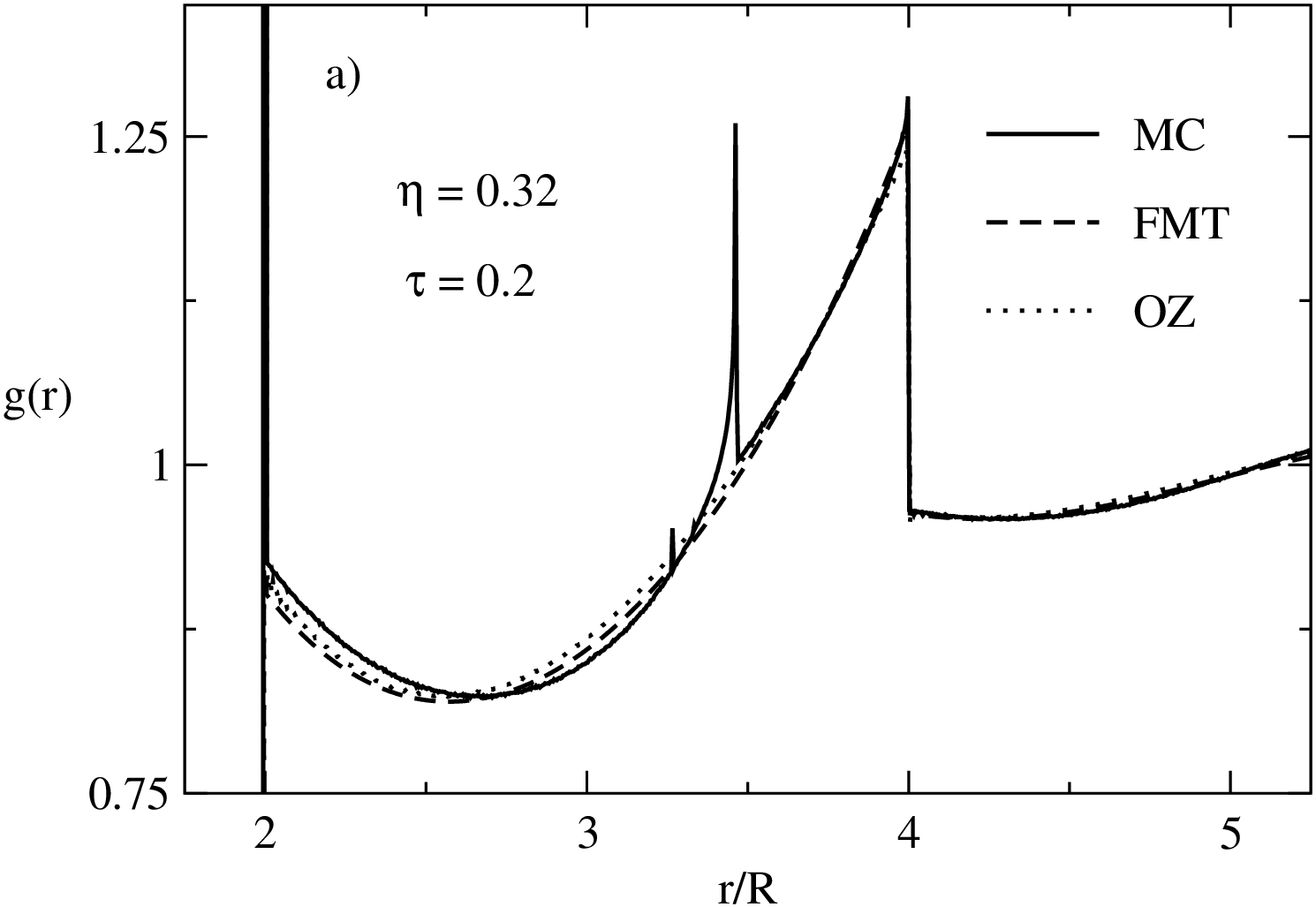} \quad \includegraphics[height=.33\textwidth]{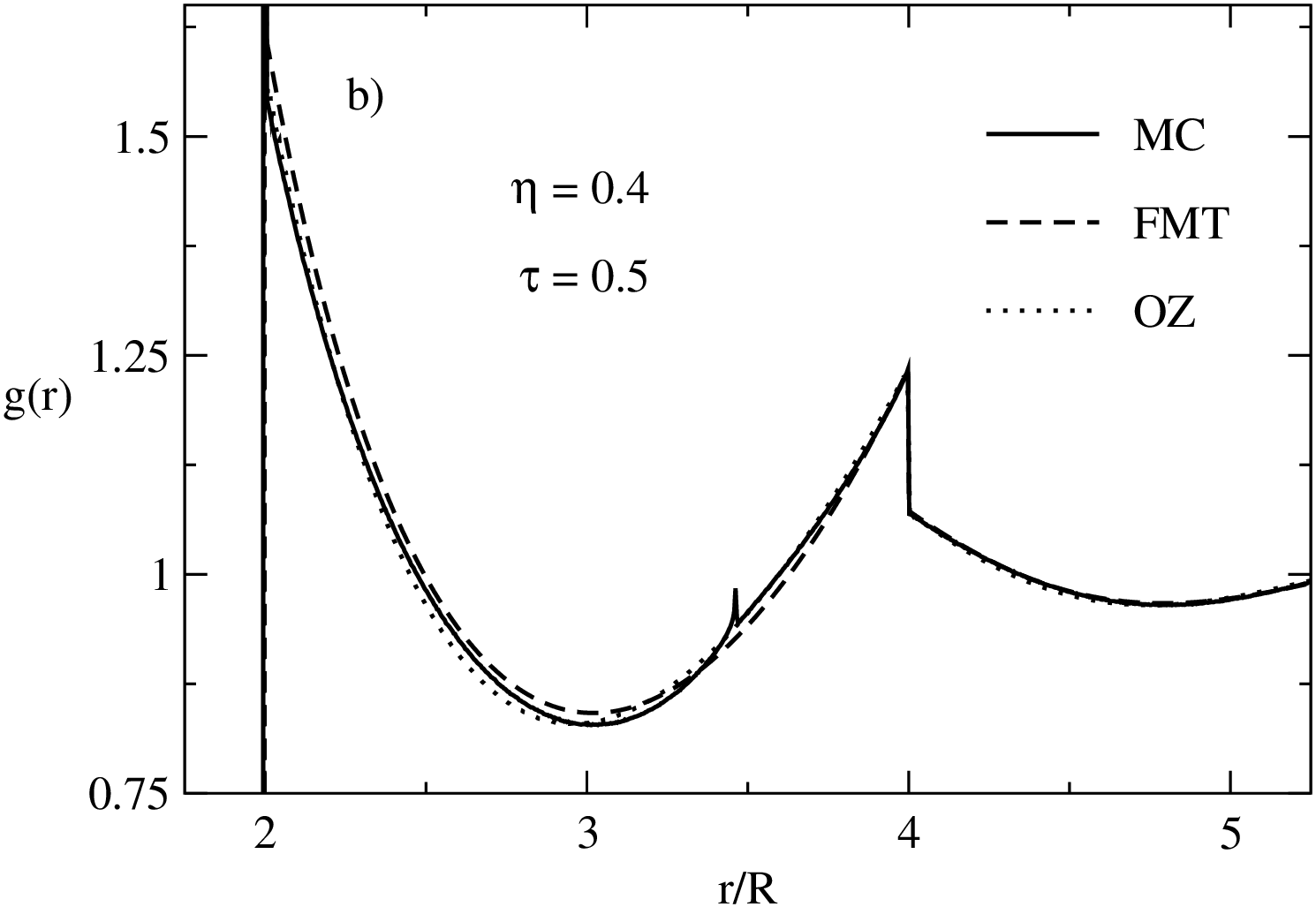}

\vspace{0.5cm}

\includegraphics[height=.33\textwidth]{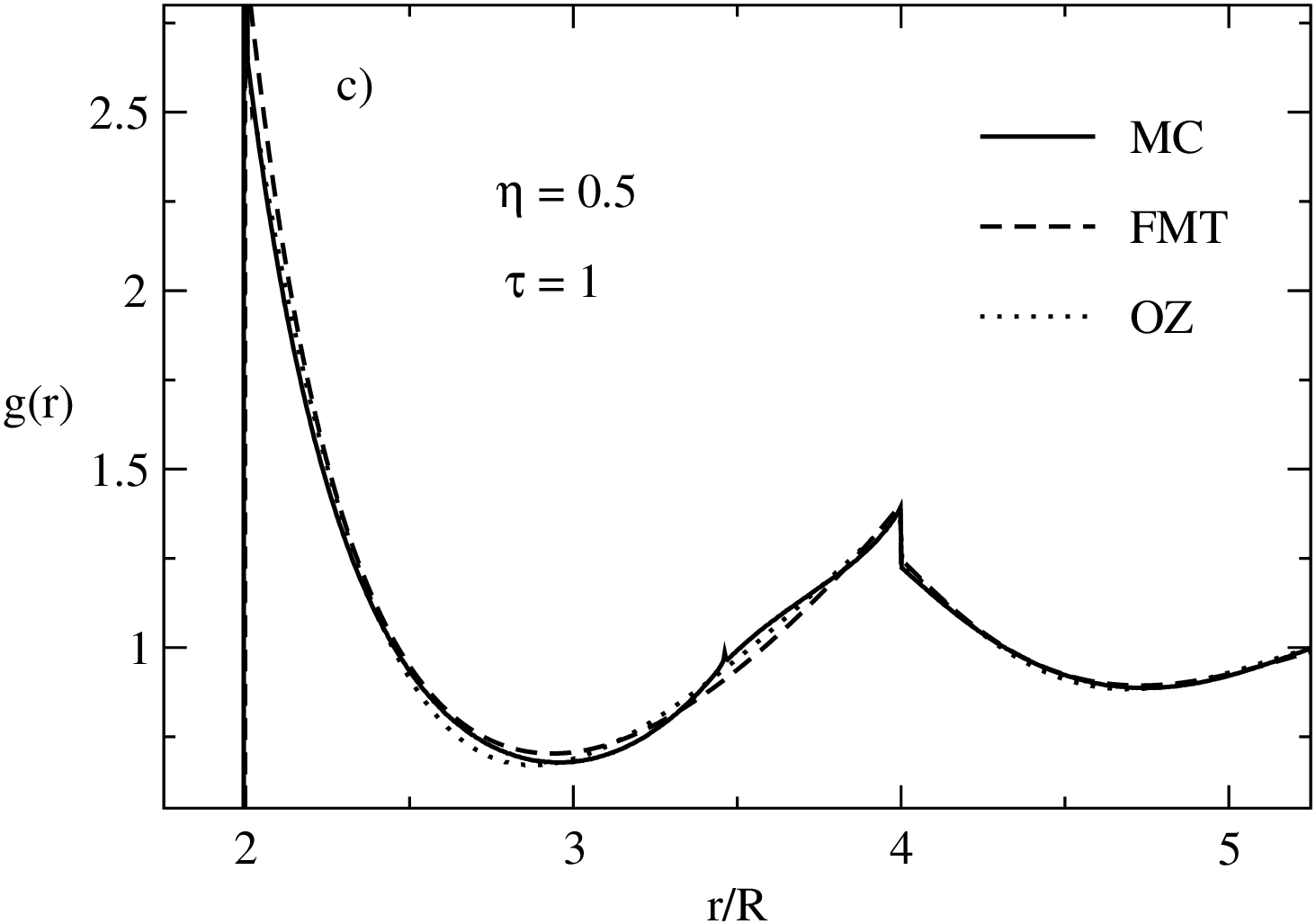}

\caption{RDFs $g(r)$ of SHS fluids with different Baxter parameters $\tau$ and packing fractions $\eta$. The full line represents the result from MC simulations \cite{MiFr04gr} and the dotted line was calculated in Ref.~\onlinecite{MiFr04gr} by using the PY result for the DCF, Eq.~\eqref{eq_GPYc} with Eq.~\eqref{eq_ysigmaPY}, in the OZ integral equation. The dashed line is the result of the numerical minimization of the density functional $\Omega$ which uses $\Phi_{\text{SHS}}$ from Eq.~\eqref{eq_defPhiSHS} and the coefficients $\phi_1^{\text{PY}}$ and $\phi_2^{\text{PY}}$ from Eqs.~\eqref{eq_phi1PY} and \eqref{eq_phi2PY}. a) $\tau = 0.2$ and $\eta = 0.32$. b) $\tau = 0.5$ and $\eta = 0.4$. c) $\tau = 1$ and $\eta = 0.5$.}

\label{fig_gr}

\end{center}

\end{figure}

In Fig.~\ref{fig_gr} we compare the FMT result for $g(r)$ from the full minimization of $\Omega$ to the results obtained (a) by using $c_{\text{PY}}$ in the OZ equation and (b) from MC simulations. Both the simulation data and the OZ results are taken from Ref.~\onlinecite{MiFr04gr}. For the three different choices of $\tau$ and $\eta$ the FMT and OZ results are consistent with each other and both agree well with the simulations. Deviations of the FMT results from the simulations occur mostly close to contact where the theory either underestimates (Fig.~\ref{fig_gr}a) or overestimates (Figs.~\ref{fig_gr}b and c) the simulation data. Hence, there does not appear to be a systematic deviation and the good agreement of the two theoretical results with the simulations confirms the validity of the new functional.

\subsection{Results for the inhomogeneous fluid}

\subsubsection{Density profiles}

\begin{figure}[t]

\begin{center}

\includegraphics[height=.32\textwidth]{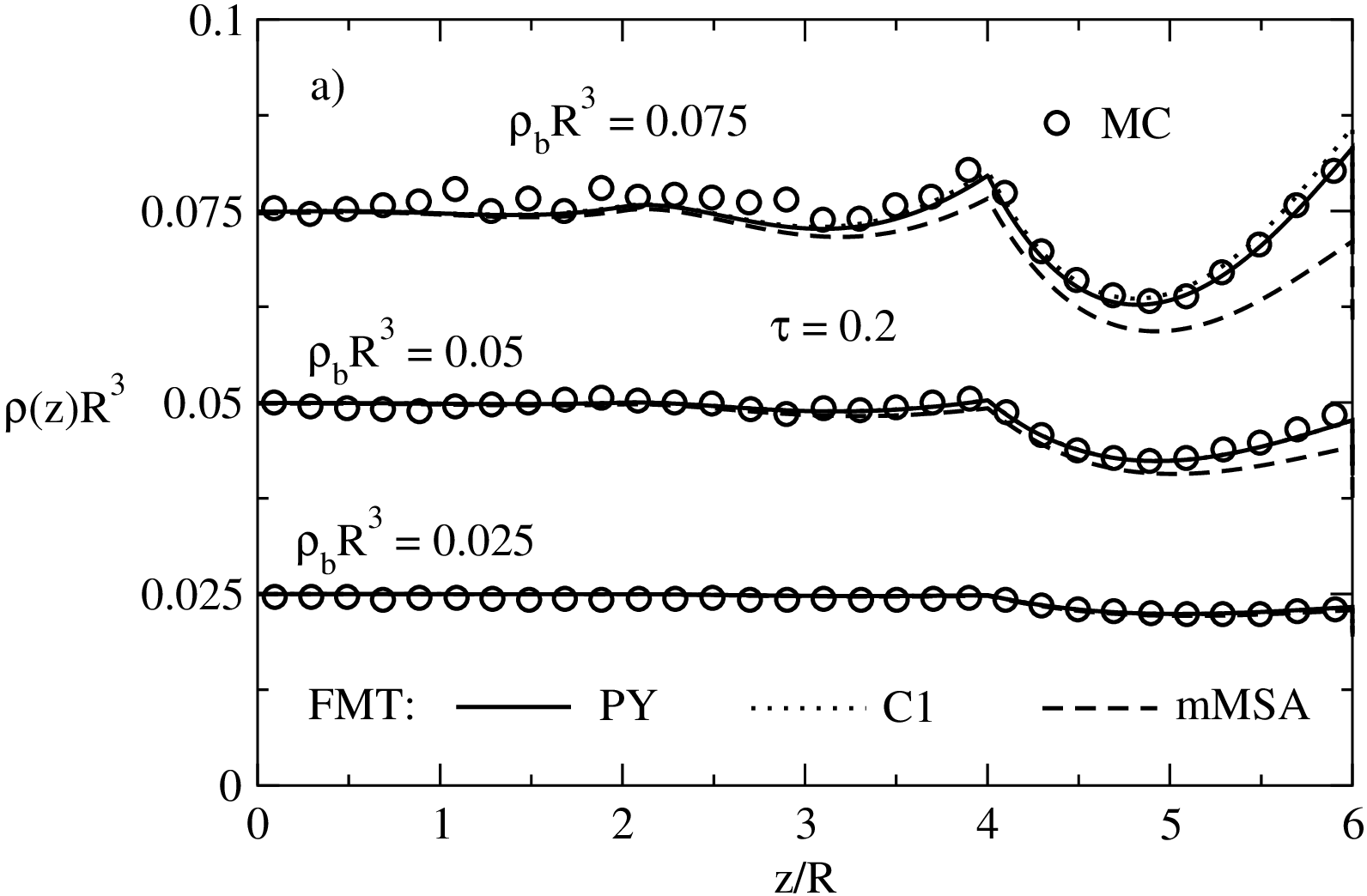} \quad \includegraphics[height=.32\textwidth]{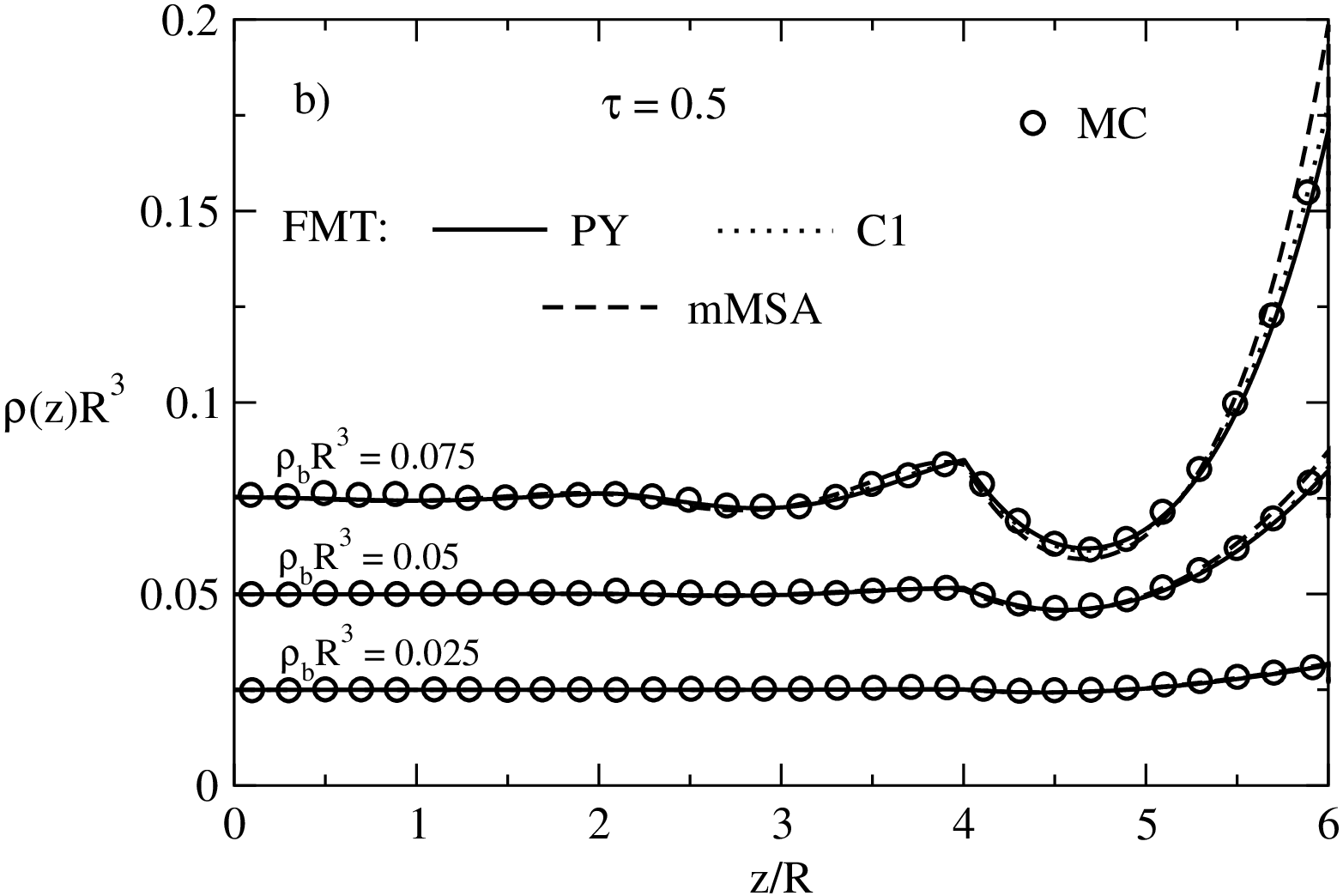}

\caption{Density profiles of a SHS fluid confined between two parallel hard walls at a distance $14 R$. The midplane is located at $z=0$. Simulation data (open circles) are taken from Ref.~\onlinecite{JaBr94}. The curves are the results from the DFT presented in Sec.~\ref{sec_FMTSHS}. The different bulk densities $\rho_b$ of the fluid are given in the figure. a) Baxter's parameter is $\tau=0.2$. b) Baxter's parameter is $\tau=0.5$.}

\label{fig_proL14tau}

\end{center}

\end{figure}

A standard approach to test a DFT's ability to describe a fluid in an external potential which creates an inhomogeneity is to consider the fluid in contact with a planar hard wall. Due to the underlying symmetry the required numerical minimization of the functional $\Omega$ is effectively a 1D computation and hence results are relatively easy to obtain. The formulas for the weighted densities in the planar case can be obtained by considering Eqs.~\eqref{eq_n3spherical} to \eqref{eq_n0spherical} in the limit $r\to\infty$. We compute density profiles for two different values of $\tau$ and three different bulk densities $\rho_b$ using the FMTs based on the PY, mMSA, and C1 closures for a fluid that is confined between two parallel walls at a distance of $14 R$. In Fig.~\ref{fig_proL14tau} the results are compared with data from canonical MC simulations taken from Ref.~\onlinecite{JaBr94}. The PY based FMT provides an excellent agreement with the simulations and the C1 based FMT performs nearly as well as the former. The mMSA based FMT shows deviations from the simulations which increase with decreasing $\tau$ and increasing $\rho_b$.  Taking into account that (a) the mMSA FMT approximates the cavity function at contact $y_{\sigma}(\eta)$ by $y_{\sigma}(0)$ and (b) the C1 FMT incorporates the density dependence of $y_{\sigma}(\eta)$ only to lowest order in $\eta$, we conclude that the quality of the underlying approximation for $y_{\sigma}(\eta)$ is not the crucial factor that determines the quality of the FMT. Indeed, the key ingredient appears to be the structure of the DCF $c_{\text{GPY}}(r)$ [Eq.~\eqref{eq_GPYc}] that follows from the GPY closure relation. The quality of the FMT results is clearly superior to the results obtained from the OZ equation based PY singlet approximation \cite{JaBr93, JaBr94} or the DFT by Choudhury and Ghosh \cite{ChGh97}, both of which deviate significantly from simulations especially close to contact with the wall. Higher order DFTs \cite{HwKi98, YoKi98,Zh01} require a fitting parameter in order to match the correct contact value or equation of state. In contrast, the fact that the FMT respects the contact theorem \cite{He89}, i.e. $\beta p = \rho_c$, {\em without} an adjustable parameter, insures that the density profile is nicely described by the present theory even close to the wall.

\subsubsection{Solvation force}

\begin{figure}[t]

\begin{center}

\includegraphics[height=.32\textwidth]{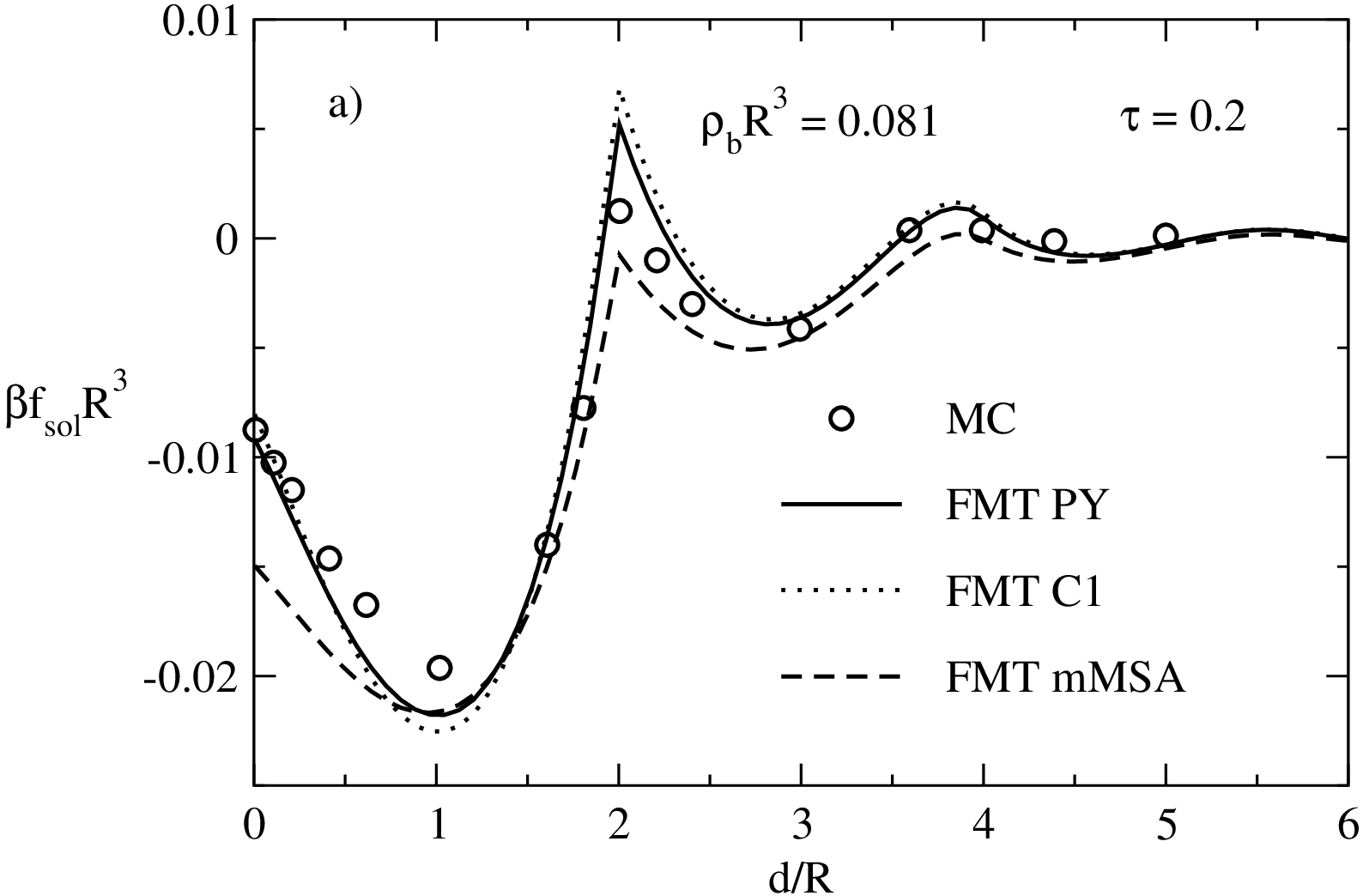} \quad \includegraphics[height=.32\textwidth]{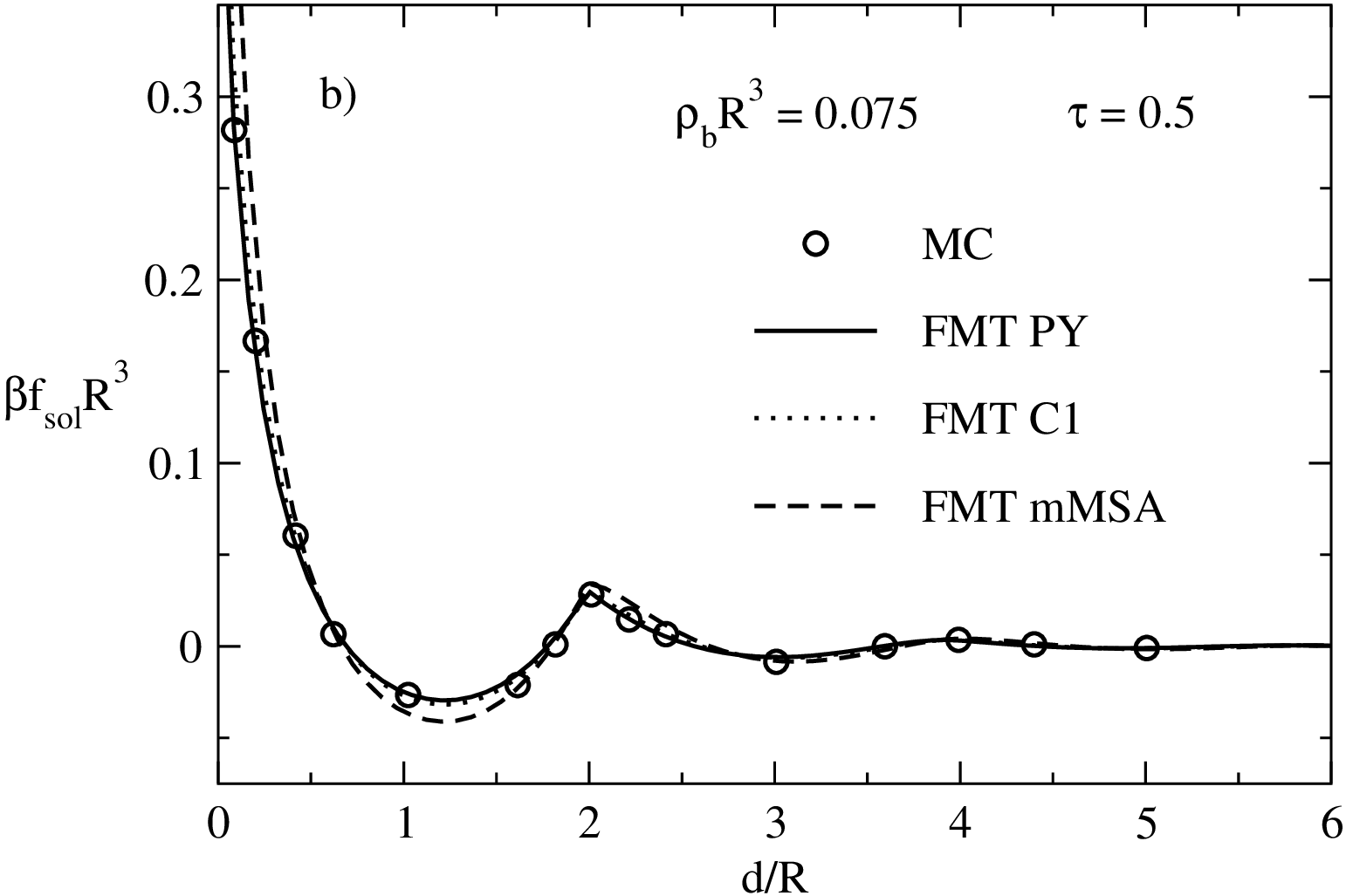}

\caption{Solvation force $f_{\text{sol}}$ between two hard plates as obtained in simulations \cite{JaBr94} compared to the results from FMT. The solvation force is calculated from the contact values of the density profile according to $\beta f_{\text{sol}}(d) = \rho_c(d) - \rho_c(\infty)$. a) The Baxter parameter is $\tau = 0.2$ and the bulk density is $\rho_b = 0.081/R^3$. b) The Baxter parameter is $\tau=0.5$ and the bulk density is $\rho_b = 0.075/R^3$.}

\label{fig_fsoltau}

\end{center}

\end{figure}

As a final test of how well the FMT compares with simulations we consider the solvation force $f_{\text{sol}}$ between two parallel hard plates at a distance $d$. The solvation force (per unit area) is closely related to the contact value $\rho_c$ of the density profile via $\beta f_{\text{sol}}(d) = \rho_c(d) - \rho_c(\infty)$. Here, we choose to plot $f_{\text{sol}}(d)$ rather than $\rho_c(d)$. This is because the accuracy of $\rho_c(\infty)$ is assured by the underlying equation of state through the contact theorem \cite{He89}, while slight differences in the equation of state lead to different offsets for the $\rho_c(d)$ curves which obscures the comparison between the results. The solvation force for a fluid with Baxter parameter $\tau = 0.2$ and bulk density $\rho_b = 0.081/R^3$ is shown in Fig.~\ref{fig_fsoltau}a. All the FMTs perform dramatically better than the OZ equation based singlet PY results \cite{JaBr94} which are particularly poor for the contact density. As expected, the mMSA FMT is inferior to the PY and C1 FMTs. In contrast to the density profiles in Fig.~\ref{fig_proL14tau}, where the PY and C1 FMT both capture the simulation data equally well, when describing $f_{\text{sol}}$ the PY FMT appears slightly superior to the C1 FMT. This is consistent with the fact that a better approximation to $y_{\sigma}$ underlies the PY FMT. Results for the more moderate case of a fluid which has $\tau=0.5$ and $\rho_b = 0.075/R^3$ are shown in Fig.~\ref{fig_fsoltau}b. The PY and C1 FMTs show excellent agreement with the simulation data while the mMSA FMT are less accurate. The former versions of FMT clearly perform better than the DFT from Ref.~\onlinecite{Zh01} (cf.\ figure 7 therein). 

\section{Conclusion}

\label{sec_conclusion}

We have constructed a DFT for the SHS fluid which, in the spirit of Rosenfeld's FMT for the HS fluid \cite{Ro89}, is based on a set of weighted densities which can be identified from a careful inspection of the dilute HS fluid. The main finding of this work is that using the FMT set of weighted densities (from the version of Kierlik and Rosinberg \cite{KiRo90,KiRo93}), in combination with an exact relation from SPT \cite{Reiss60,Reiss77}, a density functional can be constructed which yields the PY DCF \cite{Ba68} or the DCF resulting from any other closure of the broader GPY class \cite{GaGi04}. The construction is uniquely determined by the set of weighted densities [Eq.~\eqref{eq_defwd} with Eqs.~\eqref{eq_omega3real} -- \eqref{eq_omega0real}] the SPT differential equation [Eq.~\eqref{eq_SPTdiff}] and the given closure from the GPY class.

We tested the performance of the new FMT functional for the SHS fluid by comparison with simulation data for the RDF and for density profiles of the confined fluid between two planar walls. The agreement is very good with the most significant improvement over previous theories such as the singlet PY results \cite{JaBr93, JaBr94} or the DFT by Choudhury and Ghosh \cite{ChGh97} occurring in the density profiles close to contact with the wall. This reflects the property that the FMT complies with the contact theorem \cite{He89}, while other DFTs need an adjustable parameter to obtain the correct contact value \cite{YoKi98,Zh01}. The results for the solvation force between two planar plates are also significantly better than the ones obtained with previous theories \cite{JaBr94, Zh01}.

Besides the obvious advantage of providing a very accurate description of the structure of the inhomogeneous SHS fluid, the existence of an FMT formulation for the SHS fluid has several important implications for future developments. One aspect is the extension of the theory to SHS mixtures. Given that FMT provides a simple and accurate theory for HS mixtures, this new functional provides a starting point for the study of SHS mixtures. In particular, the weighted densities are the key to the simplicity of the HS mixture FMT. However, unlike this situation, the new functional depends on the particle radius $R$ and the Baxter parameter $\tau$. Hence, a straightforward generalization of the functional with the help of weighted densities becomes impossible for mixtures in which $R$ or $\tau$ or both $R$ and $\tau$ are different. A future challenge resides in seeking an appropriate set of weighted densities for the SHS fluid which shifts the $R$ and $\tau$ dependence from the functional to the weighted densities, thereby opening the door for broad applicability to SHS mixtures. The complexion of such a set might be explored through incremental extension to simple binary mixtures some of which allow for an analytical solution within the PY closure \cite{PeSm75,CuPeSm76}. Ultimately, one might be able to treat non-spherical sticky particles as was the case for hard particles \cite{Ro94}.

An additional future development includes extending the present DFT formulation to fluids with an attraction that has non-vanishing range such as the square-well fluid. Given that the PY closure cannot be solved analytically for these systems, it seems most likely that a successful road to an FMT would include the creation of new weighted densities which average the density profiles over a region that corresponds to the range of the attractive potential. In this context, studies of SW fluids with a very short but finite range of attraction \cite{MaYuSa06} might provide a good starting point.

Finally, there is the question whether the present FMT is suited for the study of SHS crystals. Problems such as spurious divergences for peaked density distributions for HS systems lead us to believe that the SHS FMT in its present form cannot properly describe a SHS crystal. However, these shortcomings have been successfully cured using different approaches \cite{RoSchmLoeTa97, TaRo97, Ta00} within Rosenfeld's formulation of FMT \cite{Ro89}. Nonetheless, these approaches have not yet been translated to the KR framework \cite{KiRo90} which constitutes the basis of the SHS FMT. The alternative approach of formulating the present SHS FMT using Rosenfeld's set of weighted densities appears to be challenging. This follows from the observation that the $\delta$ function in the DCF [Eq.~\eqref{eq_GPYc}] can only be generated using KR's weighted density $n_1$. If the vectorial $\vec{n}_2$ in Rosenfeld's FMT, supplemented by $n_3$ and $n_2$, was used to construct $\Phi_{\text{SHS}}-\Phi_{\text{HS}}$ the DCF in Fourier space would acquire a term proportional to $\vec{\omega}_2(k) \vec{\omega}_2(k)$ [see Ref.~\onlinecite{Ro90} for the definition of $\vec{\omega}_2$]. Such a formulation is unable to yield the term proportional to $\sin kR \cos kR / k$ originating from the $\delta$-function in the DCF. This term can only be matched if the product $\omega_1(k) \omega_2(k)$, with KR's $\omega_1(k)$ from Eq.~\eqref{eq_omega1}, is used. Therefore, the extension of FMT to the SHS fluid provides an example where the equivalence of the KR and Rosenfeld formulations is no longer valid. This is an interesting and unexpected finding that illustrates the advantage of one formulation over the other depending on the task at hand.

\begin{acknowledgements}

HHG is grateful to Roland Roth for stimulating discussions and to Mark A. Miller for providing the simulation data and Percus-Yevick results for the radial distribution functions. The authors acknowledge the U.S. National Science Foundation Grant No. OPP0440841, the Department of Energy Grant No. DE-FG02-05ER15741, the Helmholtz Gemeinschaft Alliance ``Planetary Evolution and Life,'' and Yale University for generous support of this research. 

\end{acknowledgements}

\end{document}